# High-resolution 3D refractive index microscopy of multiple-scattering samples from intensity images

Shwetadwip Chowdhury[1], Michael Chen[1], Regina Eckert[1], David Ren[1], Fan Wu[2], Nicole Repina[3], and Laura Waller[1,*]

**Authors information:**
[1]Department of Electrical Engineering and Computer Sciences, University of California, Berkeley
[2]Department of Molecular and Cell Biology, University of California, Berkeley
[3]Department of Bioengineering, University of California, Berkeley
**\*Corresponding Author:** waller@berkeley.edu

## Abstract

Optical diffraction tomography (ODT) reconstructs a sample's volumetric refractive index (RI) to create high-contrast, quantitative 3D visualizations of biological samples. However, standard implementations of ODT use interferometric systems, and so are sensitive to phase instabilities, complex mechanical design, and coherent noise. Furthermore, their reconstruction framework is typically limited to weakly-scattering samples, and thus excludes a whole class of multiple-scattering samples. Here, we implement a new 3D RI microscopy technique that utilizes a computational multi-slice beam propagation method to invert the optical scattering process and reconstruct high-resolution (NA>1.0) 3D RI distributions of multiple-scattering samples. The method acquires intensity-only measurements from different illumination angles, and then solves a non-linear optimization problem to recover the sample's 3D RI distribution. We experimentally demonstrate reconstruction of samples with varying amounts of multiple scattering: a 3T3 fibroblast cell, a cluster of C. elegans embryos, and a whole C. elegans worm, with lateral and axial resolutions of ≤250 nm and ≤900 nm, respectively.

# I. Introduction

Fluorescent imaging has enabled stunning visualizations of biological processes at a variety of size scales and resolutions, for studies of gene expression, protein interactions, intracellular dynamics, etc [1-4]. However, the fluorescent techniques require exogenous biological labels, and so do not directly give endogenous information about a sample's biological structure.

Optical diffraction tomography (ODT) also targets 3D biological imaging. In contrast to fluorescent methods, ODT avoids the use of exogenous biological labels, and instead utilizes the intrinsic optical variation within a sample to reconstruct its 3D refractive-index (RI) distribution [5-11]. Hence, ODT avoids some of fluorescent imaging's main drawbacks, such as photobleaching, slow acquisition speed, low signal-to-noise (SNR) ratio, and complex sample-preparation protocol. Furthermore, RI imaging enables examination of the structural, mechanical, and biochemical properties of a sample, which are important for studies in morphology, mass, shear stiffness, and spectroscopy [9, 12-15].

Standard implementations of ODT use either a rotating sample or a scanning laser beam to capture the angle-specific scattering arising from the sample [5, 7, 16-18]. Under the assumption of weak scattering (i.e., 1st Born or Rytov approximations), 2D electric-field measurements directly yield information about the sample's 3D scattering potential [19-21]. Standard ODT reconstruction algorithms utilize the Fourier diffraction theorem to project the information contained in each electric-field measurement onto spherical shells (i.e., Ewald surfaces) in the 3D Fourier space of the sample's scattering potential [22, 23]. After sufficient 3D coverage of the sample's scattering potential has been achieved, 3D RI distributions can be reconstructed [7, 24, 25]. This reconstruction framework is efficient and fast, and has found great success in visualizing individual unlabeled cells with quantitative RI.

Unfortunately, standard ODT has two main drawbacks that limit its utility in biological research: 1) it requires sensitive electric-field measurements via laser illumination and interferometry, which associate with coherent noise, phase instabilities, and complex system-alignment protocols, and 2) the Fourier diffraction theorem relies on the sample being weakly scattering, which limits applications to predominantly individual cells. This excludes a whole class of biologically-relevant samples, such as densely packed clusters of cells or multicellular organisms, which are optically transparent but multiple-scattering.

Advances in computational phase retrieval have introduced alternative techniques that utilize optimization [26-34] for 3D sample reconstruction. Unlike the standard ODT framework, which relies on an analytical inversion of the scattering process via the Fourier diffraction theorem [7, 22] or weak-phase approximation [35, 36], such computational techniques iteratively refine an initial estimate of the 3D sample towards a solution that minimizes the difference between the observed measurements and the expected measurements, as predicted by a physical forward model. While analytical inversion for multiple-scattering is often not feasible, forward models based on multiple scattering are. Hence, for scenarios where the sample is not weakly-scattering and there exist no robust analytical inverter, optimization-based iterative solvers are a promising alternative.

Recent work by Tian et al [34] introduced a computational technique to invert multiple-scattering. Angled illuminations of the sample, incorporated into the multi-slice beam-propagation

(MSBP) forward model, were used to reconstruct the sample's 3D phase distribution from intensity measurements, without physical sample scanning. To do so, however, the reconstruction framework utilized a 2D gradient-descent protocol to iteratively invert 3D scattering. This simplification accumulates error as the axial sampling density is increased, making the framework untenable for high numerical aperture (NA) 3D RI imaging. Subsequent work by Kamilov et al [33] introduced a framework, based on MSBP and sparse regularization, to rigorously invert high-NA 3D multiple-scattering from electric-field measurements captured by a laser-based interferometer. However, such interferometer systems typically associate with mechanical instabilities and coherent artifacts, which cannot be easily accounted for in the reconstruction model. Thus, this can give rise to instabilities or over-regularization in the nonlinear iterative solver, degrading effective resolution.

In this work, we present a new 3D RI microscopy technique, also based on the multi-slice beam-propagation (MSBP) forward model, that uses intensity-only measurements to iteratively reconstruct 3D RI of multiple-scattering biological samples. Similar to the works described above, our technique illuminates the sample from different angles in order to encode 3D information into 2D measurements, then recovers the sample's 3D RI distribution. Beyond the work by Tian et al [34], we present a rigorous nonlinear optimization protocol that performs robustly at high resolutions (NA > 1.0) and enables full axial sampling of the system's diffraction limited point-spread-function. Even better resolution may be achieved due to the multiple-scattering within the sample [34, 37, 38], but is difficult to specifically characterize. To the best of our knowledge, this work presents the first demonstration of high-resolution (250 nm lateral and 900 nm axial resolution) inversion of 3D biological multiple-scattering from intensity-only measurements.

# II. Theory

## A. Multi-slice beam-propagation forward model

Beam-propagation methods generally use a set of initial conditions to calculate the electric-field at a separate location in space (or time) [39, 40]. Multi-slice beam-propagation (MSBP) methods specifically approximate 3D objects as a series of thin layers, where light propagation through the sample is modelled via sequential layer-to-layer propagation of the electric-field [33, 34, 41]. Mathematically, this can be recursively written as:

$$y_k(\mathbf{r}) = t_k(\mathbf{r}) \cdot \mathcal{P}_{\Delta z}\{y_{k-1}(\mathbf{r})\} \qquad (1)$$

where $\mathbf{r}$ denotes the 2D spatial position vector, $\{y_k(\mathbf{r}) \mid k = 1,2,\dots N\}$ denotes the electric-field at the $k$th layer of an object that is approximated with $N$ equally-spaced layers separated by distance $\Delta z$, $\left\{t_k(\mathbf{r}) = \exp\left(j\left(\frac{2\pi}{\lambda}\right)\Delta z\,(n_k(\mathbf{r}) - n_m)\right) \mid k = 1,2,\dots N\right\}$ is the object's $k$th-layer complex transmittance, $\{n_k(\mathbf{r}) \mid k = 1,2,\dots N\}$ is the object's $k$th-layer complex-valued refractive index, and $n_m$ denotes the homogenous refractive index of the surrounding media. $\mathcal{P}_{\Delta z}\{.\}$ is the mathematical operator to propagate an electric-field by distance $\Delta z$, according to the angular spectrum propagation method, i.e., $\mathcal{P}_{\Delta z}\{.\} = \mathcal{F}^{-1}\left\{\exp\left(-j\,\Delta z\left(\left(\frac{2\pi}{\lambda}\right)^2 - |\mathbf{k}|^2\right)^{1/2}\right)\mathcal{F}\{.\}\right\}$, with $\mathcal{F}\{.\}$ and $\mathcal{F}^{-1}\{.\}$ denoting the 2D Fourier and inverse Fourier transforms, respectively, and $\mathbf{k}$ denting the 2D spatial frequency vector. The boundary condition to initialize Eq. (1) is simply the incident planar electric field illuminating the sample at a particular angle, i.e., $y_0(\mathbf{r}) = \exp(j\,\mathbf{k_0}\cdot\mathbf{r})$, where $\mathbf{k_0}$ is the 2D illumination wave-vector.

The exit electric-field, $y_N(\mathbf{r})$, accounts for the accumulation of the diffraction and multiple-scattering processes that occurred during optical propagation through the sample, and contains information about the sample's 3D structure. The final electric-field and intensity distributions at the image plane are,

$$E(\mathbf{r}) = \mathcal{F}^{-1}\{p(\mathbf{k})\cdot\mathcal{F}\{\mathcal{P}_{-\hat{z}}\{y_N(\mathbf{r})\}\}\} \tag{2}$$

$$I(\mathbf{r}) = |E(\mathbf{r})|^2 \tag{3}$$

where $p(\mathbf{k})$ denotes the system's pupil function, and $\hat{z}$ refers to the distance between the plane of $y_N(\mathbf{r})$ and the plane within the object's volume at conjugate focus to the imaging plane of the system. When the imaging system is focused at the center of the object, $\hat{z} = \Delta z\,N/2$.

## B. Inverse problem formulation

Multiple measurements of the object are captured at varying illumination angles. We denote $y_0^\ell(\mathbf{r}) = \exp(j\,\mathbf{k_0^\ell}\cdot\mathbf{r})$ for $\ell = 1, 2, \ldots, L$ as a set of $L$ planar electric-fields incident onto the first layer of the sample, at various angles set by their wave-vectors $\mathbf{k_0^\ell}$. The corresponding intensity measurements are denoted by $I^\ell(\mathbf{r})$.

The reconstruction framework can be formulated as a least-squares minimization that seeks to estimate the sample's 3D RI by minimizing the difference between the measured amplitude (square-root of intensity [42]) and those expected via the forward model,

$$\hat{n}(\mathbf{r}_{3D}) = \underset{n(\mathbf{r}_{3D})}{\arg\min}\sum_{\ell=1}^{L}\sum_{\mathbf{r}}\left|\sqrt{I^\ell(\mathbf{r})} - \left|\mathcal{G}^\ell\{n(\mathbf{r}_{3D})\}\right|\right|^2 \tag{4}$$

Here, $\mathbf{r}_{3D} = \langle\mathbf{r}, k\rangle$ is a 3D spatial position vector, such that $n(\mathbf{r}_{3D}) = n_k(\mathbf{r}), k = 1,2,3,\ldots,N$. The non-linear operator $\mathcal{G}^\ell\{.\}$ denotes the $\mathbb{C}^3 \to \mathbb{C}^2$ forward model operation that predicts the 2D electric-field distribution measured at the camera after illuminating $n(\mathbf{r}_{3D})$ with the incident electric-field $y_0^\ell(\mathbf{r})$, as described by Eqs (1) and (2).

## C. Reconstruction framework

Our proposed framework to minimize Eq (4) and reconstruct $\hat{n}(\mathbf{r}_{3D})$ draws from previous MSBP works by Kamilov et al [33] and Tian et al [34], and utilizes an iterative scheme to interleave back-propagation and gradient-descent steps into each iteration. We describe below the steps taken for the iterative reconstruction.

1. Initialize a $N$-layer reconstruction volume with constant value $n_m$. This will serve as the initial estimate of $n(\boldsymbol{r}_{3D})$. Then, initialize the iteration index at this step to $d = 0$.

2. This step signals the start of a new iteration. Increment the iteration index, $d \leftarrow d + 1$, and initialize the per-iteration cost function, $c(d) = 0$

3. Randomly choose (without replacement) an angular-specific illumination electric-field $y_0^\ell(\boldsymbol{r}) = \exp(j\, \boldsymbol{k}_0^\ell \cdot \boldsymbol{r})$, with associated raw intensity measurement $I^\ell(\boldsymbol{r})$, from the complete set of $\ell = 1, 2, \ldots, L$.

4. For the chosen illumination field $y_0^\ell(\boldsymbol{r})$, use Eqs. (1) and (4) to recursively compute and store the electric-field at each layer of the reconstruction volume $\{y_k^\ell(\boldsymbol{r}) \mid k = 1, 2, \ldots, N\}$, and the estimate of the final intensity measurement as predicted by the forward model, $\mathcal{G}^\ell\{n(\boldsymbol{r}_{3D})\}$.

5. Increment the cost function for the current iteration, $c(d) \leftarrow c(d) + \sum_r \left| \sqrt{I^\ell(\boldsymbol{r})} - |\mathcal{G}^\ell\{n(\boldsymbol{r}_{3D})\}| \right|^2$

6. Initialize a residual term denoted by $q_{N+1}^\ell(\boldsymbol{r})$. The variable $q_0(\boldsymbol{r})$ below is used only for notational simplicity.

$$q_0(\boldsymbol{r}) = \exp(j\, \angle \mathcal{G}^\ell\{n(\boldsymbol{r}_{3D})\}) \cdot \left( |\mathcal{G}^\ell\{n(\boldsymbol{r}_{3D})\}| - \sqrt{I^\ell(\boldsymbol{r})} \right) \quad (5)$$

$$q_{N+1}^\ell = \mathcal{P}_{\hat{z}}\left\{ \mathcal{F}^{-1}\left\{ \overline{p(\boldsymbol{k})} \cdot \mathcal{F}\{q_0(\boldsymbol{r})\} \right\} \right\}$$

7. For each layer of the reconstruction volume occupied by the object, compute the back-propagation term $s_k^\ell(\boldsymbol{r})$ by recursively propagating backwards (i.e., $k = N, (N-1), \ldots, 2, 1$),

$$s_k^\ell(\boldsymbol{r}) = \left( -j \frac{2\pi \Delta z}{\lambda} \right) \cdot \overline{t_k(\boldsymbol{r})} \cdot \overline{\mathcal{P}_{\Delta z}\{y_{k-1}^\ell(\boldsymbol{r})\}} \cdot q_{k+1}^\ell(\boldsymbol{r}) \quad (6)$$

$$q_k^\ell(\boldsymbol{r}) = \mathcal{P}_{-\Delta z}\left\{ \overline{t_k^\ell(\boldsymbol{r})} \cdot q_{k+1}^\ell(\boldsymbol{r}) \right\} \quad (7)$$

$$n_k(\boldsymbol{r}) \leftarrow n_k(\boldsymbol{r}) - \alpha \cdot s_k^\ell(\boldsymbol{r}) \quad (8)$$

Note that Eq. (8) illustrates the gradient-descent step for the back-propagation process, where $\alpha$ is often a manually-tuned parameter to adjust the step-size.

8. Repeats steps (3)-(7) for each illumination angle $\ell = 1, 2, ..., L$, to incrementally refine $\{n_k(\boldsymbol{r}) \mid k = 1,2, ... N\}$ via diversity of illumination angles. After one round through all the illumination angles is complete, consolidate all the RI layers into a single 3D RI volume, $\{n_k(\boldsymbol{r}) \mid k = 1,2, ... N\} \Rightarrow n(\boldsymbol{r}_{3D})$.

9. Conduct a 3D total-variation (TV) regularization process on $n(\boldsymbol{r}_{3D})$ to stabilize the iterative convergence in the presence of physical effects in the imaging system that are unaccounted for in the MSBP forward model, such as camera noise, coherent source noise, optical aberrations, etc. The strength of regularization is set by the parameter $\beta$.

$$n(\boldsymbol{r}_{3D}) \leftarrow \text{prox}\{n(\boldsymbol{r}_{3D}), \beta\} \tag{9}$$

where the operator $\text{prox}\{f(\boldsymbol{r}_{3D}), \gamma\}$ is generally defined for some 3D function $f(\boldsymbol{r}_{3D})$ and parameter $\gamma$ as,

$$\text{prox}\{f(\boldsymbol{r}_{3D}), \gamma\} = \underset{g(\boldsymbol{r}_{3D})}{\arg\min} \left\{ \frac{1}{2} \| f(\boldsymbol{r}_{3D}) - g(\boldsymbol{r}_{3D}) \|_{\ell^2}^2 + \gamma \, \text{TV}[g(\boldsymbol{r}_{3D})] \right\} \tag{10}$$

where $\text{TV}[.]$ denotes the 3D total-variation norm. The final $n(\boldsymbol{r}_{3D})$, after being output by the regularizer, now becomes the current iterative estimate of the object's 3D RI.

10. Repeat steps (2)-(9) to continue the iterative process, until convergence is reached. This point can be identified when the iterative cost function $c(d)$ levels out with respect to iteration index $d$.

# III. Experimental Methods

## A. Optical system design

Our experimental setup is shown in Fig. 1. Green ($\lambda$ = 532 nm center wavelength) LED light (Thorlabs M530F2) is fiber-coupled into multimode fiber with a 50 µm core diameter (Thorlabs M14L). Although the LED source is not natively coherent, the fiber acts as a spatial filter to sufficiently increase the LED source's spatial coherence, while still avoiding speckle and other coherent artifacts. The output light from the fiber is then collimated and directed to a mirror mounted on a motorized kinematic mount (M, Thorlabs KS1-Z8), for programmable tip/tilt capabilities. The plane of the mirror is conjugated to the biological sample through a 4f-system consisting of an achromatic doublet and a condenser objective (L1 → OBJ), such that tip/tilt of M directly results in scanning the illumination angle at the sample. The light exiting the sample is then collected through a second 4f-system, consisting of an imaging objective and an achromatic doublet (OBJ → L2). In the system, the condenser and imaging objectives (OBJ) were identical high-NA microscope objectives (Nikon, CFI Plan Apo Lambda 100x, NA 1.45). Finally, a third 4f-system was used to de-magnify the transmission output, before imaging onto a 20-megapixel sensor (CMOS, FLIR BFS-U3-200S6M). An adjustable iris (Ir, Thorlabs ID15) in the Fourier plane enables variable tuning of the system's NA. To avoid aberrations, we limit the NA of the detection system to NA = 1.1. This corresponds to a theoretical lateral and axial resolution of 240 nm and 890 nm, respectively, in the case of a weakly-scattering sample.

## B. Data acquisition

The motorized kinematic mount for angle scanning was controlled via MATLAB, and was programmed to scan the illumination through 120 angles along a spiral trajectory that fills the pupil (Fig. 1(a)). As each scan point is reached, an Arduino board sends a trigger pulse to the sensor. Each image was captured with ~40 ms of integration time, resulting in a total acquisition time of

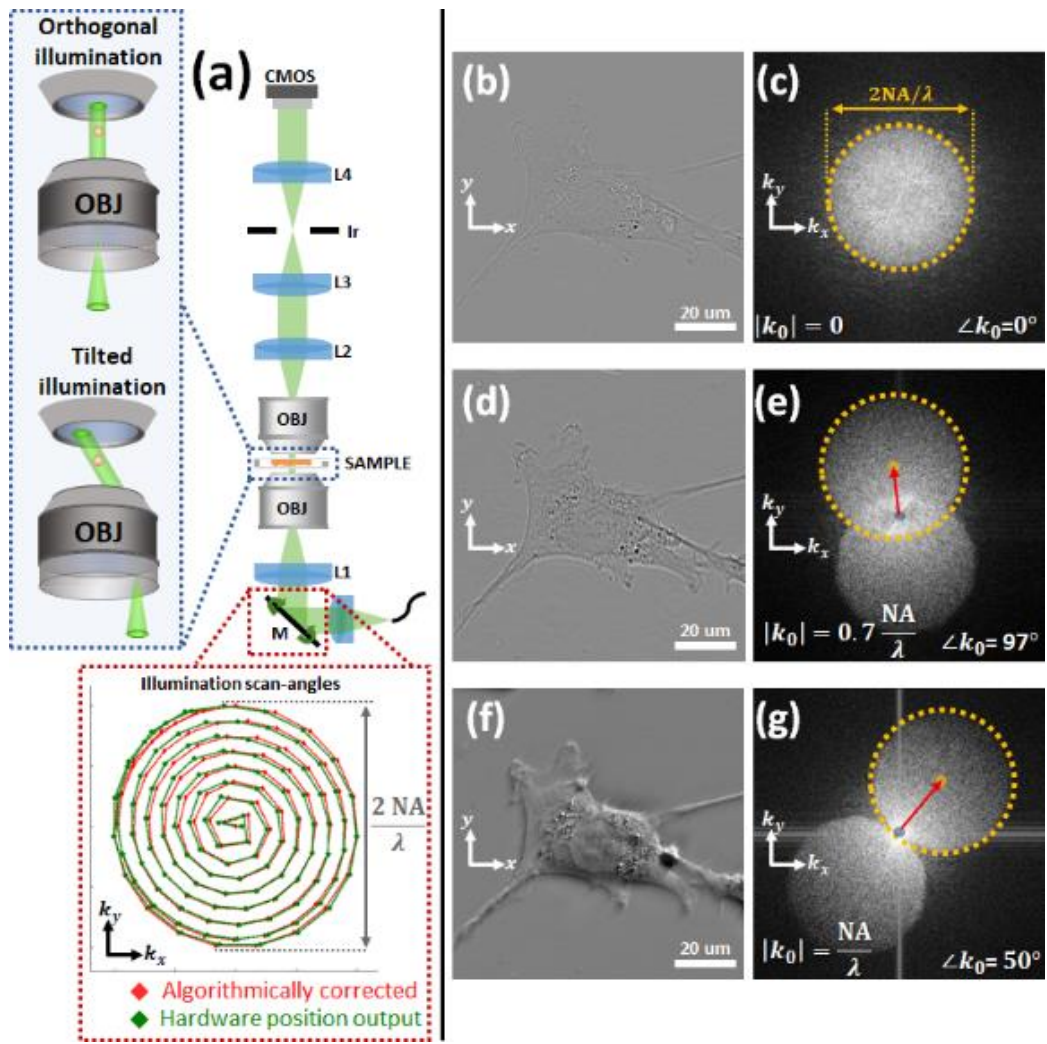

**Fig. 1. (a)** Optical system design with illumination angle trajectories, comparing the raw position values outputted from the kinematic tilt mirror to that after algorithmic self-calibration. **(b,d,f)** Raw intensity acquisitions and **(c,e,g)** amplitudes of associated Fourier transforms, after illuminating the sample with varying angles.

4.8 seconds. Reconstruction was performed with MATLAB running on a desktop computer equipped with Intel(R) Core(TM) i7-5960X CPU @ 3.00GHz 64GB RAM CPU and NVidia's GeForce GTX TitanX GPU. Total reconstruction time to complete 50 iterations of a 1200×1200×120 voxel volume was 20.3 hours.

Fig. 1(b,d,f) and Fig. 1(c,e,g) show the raw intensity measurements and associated Fourier magnitude, respectively, of a 3T3 fibroblast cell under different illumination angles. The Fourier transforms demonstrate that the spatial-frequency information is mostly contained within two circular regions in Fourier space, symmetrically positioned around the origin. This is expected for intensity-based imaging of weakly-scattering objects with off-axis illumination [35, 43] (see Supplementary Information for details).

## C. Calibration of angular illuminations

Our reconstruction requires precise knowledge of the illumination angles at the sample. Though illumination angles are outputted by the motorized kinematic mount, we experimentally found that such values did not have sufficient accuracy for satisfactory 3D RI reconstruction (likely due to system imperfections, sample-induced changes and misalignments). Thus, we instead use post-acquisition algorithmic angle-calibration as our means for precise angle measurement. We leverage previous developments in self-calibration to precisely estimate the incident illumination angles from the raw measurements [43]. Fig. 1(c,e,g) shows that the illumination wavevectors $\boldsymbol{k}_0^\ell$ can be estimated from the acquisitions' Fourier transforms, since the center of the circles (illustrated with red arrows) correspond to the illumination angle (and its conjugate). For comparison, Fig. 1(a) shows the illumination angle values outputted by the hardware alongside those estimated by the self-calibration algorithm.

# IV. Experimental results

We experimentally demonstrate our method for 3D RI reconstruction of calibration objects and biological samples with varying amounts of multiple scattering. We first image polystyrene beads for validation, and then image a fibroblast cell, as an example of a weakly-scattering sample. Finally, we look at multiple-scattering samples - C. elegans embryos and whole worm. The Supplementary Information provides a comparison between the reconstruction quality obtained by our MSBP framework and the 1st Born reconstruction framework [32].

## A. Polystyrene microspheres

Fig. 2 shows the 3D RI reconstruction of a conjoined pair of polystyrene spheres, mounted in immersion oil ($n(\lambda) = 1.552$ at $\lambda = 532$ nm). The reconstruction used iterative step-size and regularization parameter set to $\alpha = 6 \times 10^{-4}$ and $\beta = 4 \times 10^{-4}$, respectively. Fig. 2(a) shows a lateral slice through the center of the reconstruction volume. Fig. 2(b) shows an axial slice taken across the horizontal dashed-white line in Fig. 2(a). And Fig. 2(c) shows a 3D rendered visualization of the reconstruction volume, clearly capturing the spherical geometry of the microspheres. A line-profile across a single microsphere (Fig. 2(d)) shows that the experimentally measured RI is in good agreement with the expected shape and RI of polystyrene ($n_s(\lambda) = 1.598$ at $\lambda = 532$ nm) [44].

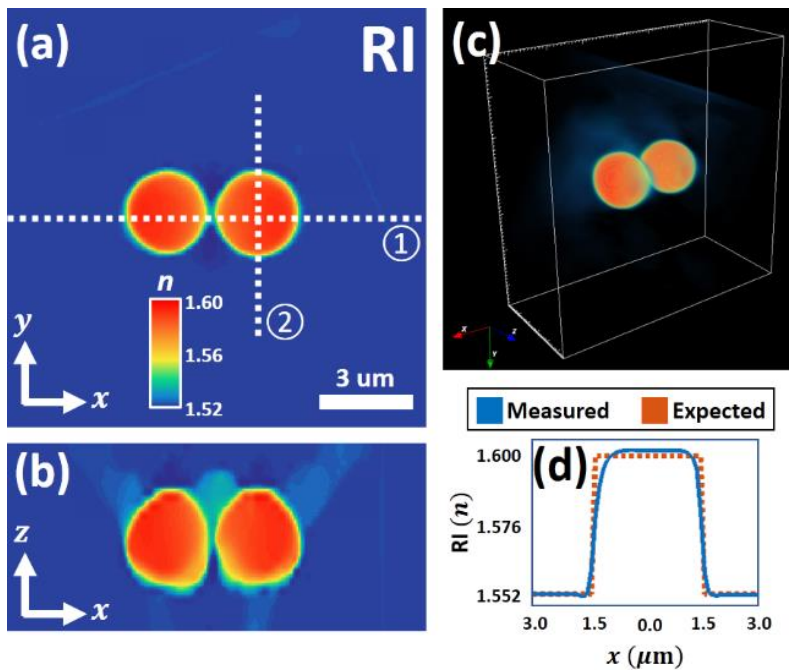

**Fig. 2.** 3D reconstruction of refractive index (RI) for two 3 um diameter polystyrene microspheres (*n*=1.598) in index-matched oil (*n*=1.552). **(a)** Lateral (x-y) slice, **(b)** axial (x-z) slice, taken along the horizontal dashed line in (a), and **(c)** 3D rendering of the reconstructed microspheres. **(d)** Cross-cut of the refractive index along the vertical dashed line in (a) matches well with expected values from the sample.

## B. Fibroblast cells

We next demonstrate RI reconstruction (Supplementary videos 1-3, 700×700×40 voxels, $\alpha = 4 \times 10^{-4}$, $\beta = 1 \times 10^{-5}$) of a weakly-scattering fibroblast (3T3) cell, using our MSBP model. Fig. 3 shows both grayscale and colored (for easier RI visualization after halo-removal [45]) lateral cross-sections at different depths through the reconstruction volume ($z = 0.0, 1.05,$ and $2.10$ um, where $z = 0.0$ um designates the attachment interface of the fibroblast cell to the coverslip). As adherent, migratory cells, fibroblast cells are known to form dynamic membrane protrusions [46]. These protrusions are clearly visualized in the reconstructed volume, with long filopodial extensions (indicated by yellow arrows in Fig. 3(c)) and broader lamellipodia and membrane ruffling at cell edges (red arrows in Fig. 3(a,c)). Intracellular compartments also have strong contrast, notably the cell nucleus with the surrounding nuclear envelope and internal nucleoli. We show a zoom-in of this region in Fig. 3(e) to highlight the nuclear envelope (red arrows), internal nucleoli (yellow arrows), and outer region of optically-dense cellular structure.

The dense fibrous network within the cell is clearly visualized and exhibits RI of 1.335-1.34 for the individual fiber tendrils. Intracellular lipids and nucleoli consistently demonstrate RI > 1.35 and ~1.335, respectively, and the internuclear space shows a RI of ~1.33, matching closely the surrounding media. Nuclear RI being lower than that of the general cell cytoplasm has also been observed in previous works for various cell lines [7, 47, 48]. Fig. 3(g) shows a tomographic rendering for easy identification of the 3D cell morphology.

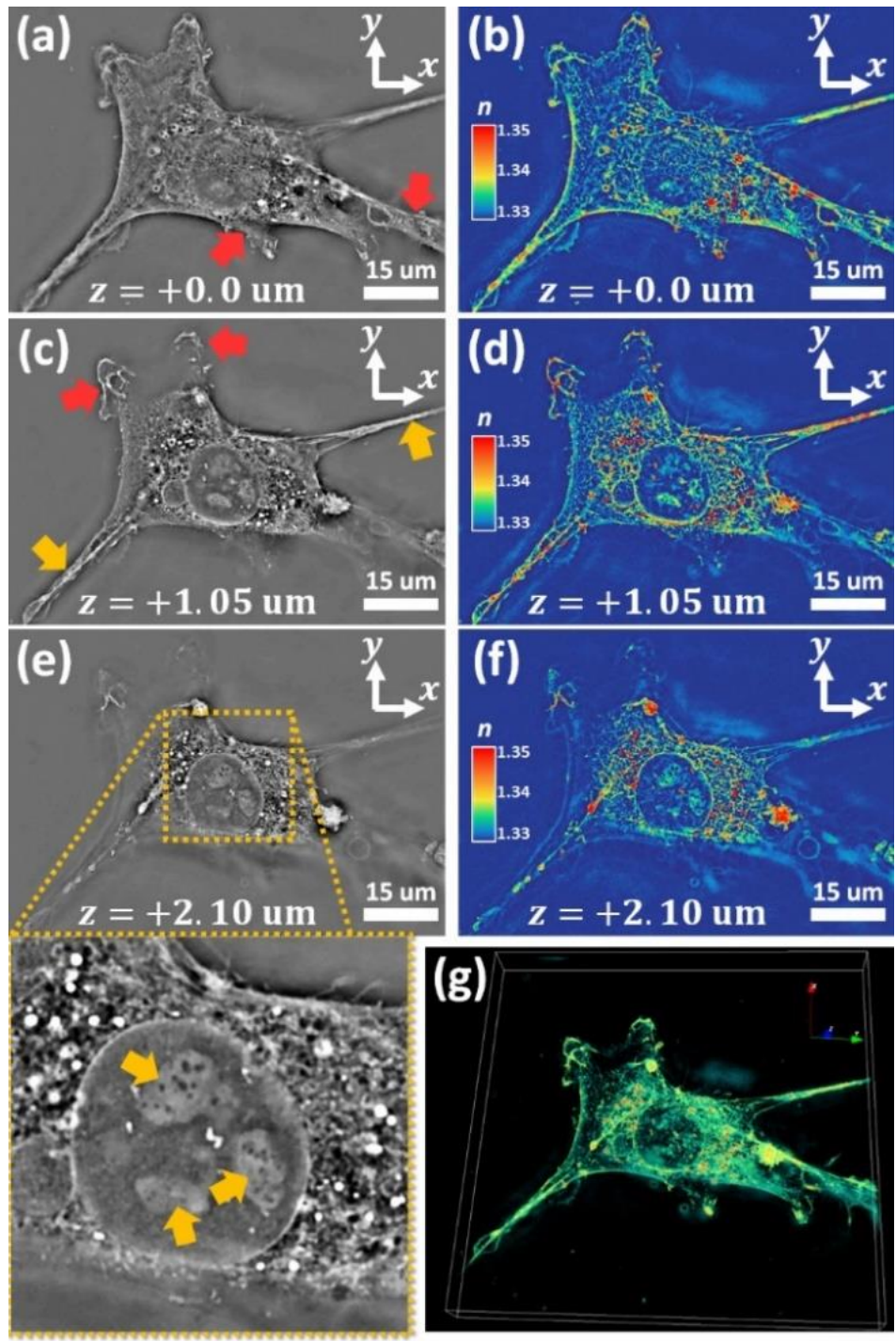

**Fig. 3.** Lateral cross-sections (gray-scale and RGB color-coded) through the 3D RI reconstruction volume at axial positions of **(a,b)** $z = 0.0$ um, **(c,d)** $z = 1.05$ um, and **(e,f)** $z = 2.10$ um. **(g)** 3D rendering of the 3T3 cell RI.

## C. *Caenorhabditis elegans* embryos

Next, we image Caenorhabditis elegans (C. elegans) embryos captured at varying developmental stages [49]. Unlike the 3T3 cell, embryos are multicellular clusters, and are thus more condensed than any one individual cell. Thus, we expect such samples to exhibit more multiple scattering. In Fig. 4(a-c), we show the raw intensity measurements and associated Fourier amplitudes of the embryo clusters under three different illumination angles. Immediately, we see from Fig. 4(a) that the condensed internal structure within the C. elegans embryos results in significant intensity contrast even under orthogonal plane-wave illumination. This contrasts with the observations of the 3T3 fibroblast cell sample, which had virtually no intensity contrast under orthogonal illumination (see Fig. 1(b)). This suggests that though the fibroblast cell can be modelled as weakly-scattering, C. elegans embryos cannot.

More rigorous evidence of multiple scattering comes from the Fourier transforms of the intensity acquisitions. As mentioned earlier, and described in more detail in the Supplementary Information, off-axis illumination of a weakly-scattering sample results in intensity acquisitions where the spatial-frequencies lie within two circular regions in Fourier space. While the intensity spatial-frequency content of the 3T3 fibroblast cell (Fig. 1(c,e,g)) was mainly contained in these circular regions, the C. elegans embryos show significant spatial-frequency content outside these regions, as indicated by yellow arrows in Fig. 4(a-c). This provides further evidence of multiple scattering.

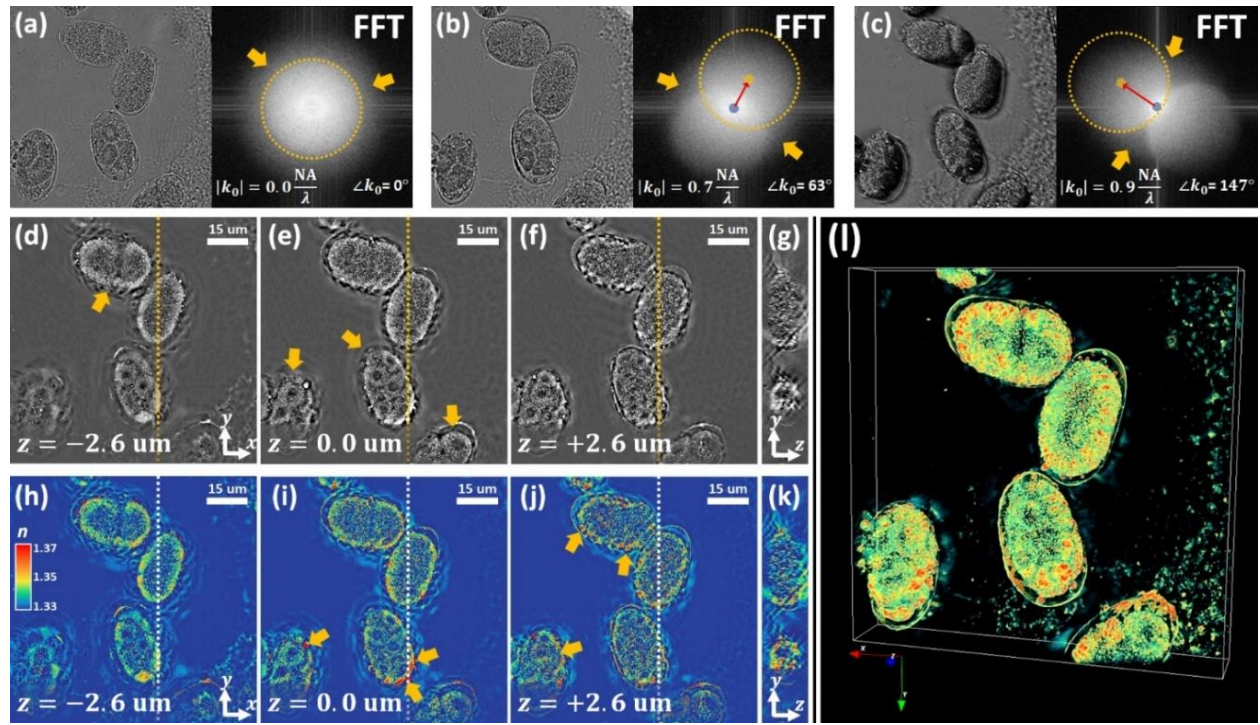

**Fig. 4.** 3D RI reconstruction of a *C. elegans* embryo cluster is presented. **(a-c)** Raw intensity-acquisitions at varying illumination angles are shown with corresponding Fourier transforms. **(d,e,f)** Lateral cross-sectional planes through the 3D reconstruction volume are shown, at axial positions $z = -2.6, 0, +2.6$ um, respectively. **(g)** Axial cross-sectional plane, taken from the reconstruction volume at the location indicated by dashed yellow in (d-f) is shown. **(h,i,j,k)** Lateral and axial cross-sectional views shown in (d,e,f,g) are redisplayed in RGB-color to facilitate easy visual inspection of refractive index. Color-bar to decode RI from color is located in (h). **(l)** 3D tomographic rendering of the embryo cluster is shown.

We show lateral slices from the 3D RI reconstruction (Supplementary videos 4-6, 1100×1100×120 voxels, $\alpha = 6 \times 10^{-4}$ and $\beta = 3.5 \times 10^{-4}$) of the embryo sample in both gray-scale (Fig. 4(d-f)) and color-scale Fig. 4(h-j), for three different depths ($z = -2.6, 0,$ and $+2.6$ um). Fig. 4(g,k) show axial cross-sections, and Fig. 4(l) shows a 3D rendering. The features visualized in this 3D RI reconstruction fit well with their developmental stages, and individual cellular compartments can be identified. Yellow arrows in Fig. 4(d) and Fig. 4(e) indicate embryos visualized during their comma stage and general 8-cell to 26- cell stages, respectively. The axial cross-section in Fig. 4(g) demonstrates in 3D the embryos' globular nature and heterogenous composition, with RI values ranging from 1.33 to above 1.37, and shows a bulk RI increase towards the periphery of the embryos. Distinct eggshells are visualized encapsulating all embryos, and demonstrate a moderately high RI ~1.35, which matches their known lipid-rich composition. Moreover, the lipoprotein complexes containing lipid and yolk proteins, localized between the embryo and eggshell, show the highest RI > 1.375 (indicated with yellow arrows in Fig. 4(i,j)).

## D. Whole *Caenorhabditis elegans* worms

Our final experiment generates a high-resolution 3D RI reconstruction of a whole adult hermaphrodite C. elegans worm (Supplementary videos 7-10, 1914×10408×118 voxels, $\alpha = 4 \times 10^{-4}$, $\beta = 4 \times 10^{-4}$). Because the imaging objective's field-of-view did not fit the whole worm, the final reconstruction volume was stitched together from 14 individually-reconstructed patches (details in Supplementary Information). We note that unlike the embryos sample, the spatial-frequencies contained within the intensity measurements of the worm (shown in Supplementary Information) do not contain any discernable symmetric circular regions. This indicates that the multiple scattering in the worm sample dominates over the single-scattering, and that the whole C. elegans worm is even more multiply-scattering than the embryos, as expected.

In Fig. 5(a,b), we show two lateral slices through the reconstruction volume at axial positions of $z = 0.0$ and $-4.9$ um, respectively. The major components of the reproductive and digestive systems are clearly identified, and include the pharynx, intestine, gonads, spermathecal, and fertilized eggs [50]. Fig. 5(c) presents an axial cross-section of the pharynx, and clearly demonstrates the interior three-fold rotationally symmetric organization of the muscles and marginal cells around the pharyngeal lumen. Visualizing this morphology typically requires a heavily invasive preparation protocol (cross-sectional slicing through the C. elegans pharynx followed by electron microscopy) [51]. Fig. 5(d,e) present a zoom-in of the worm's mouth and pharyngeal tip at axial positions of $z = +2.42$ and $+5.19$ um. Yellow arrows indicate E. coli bacteria, a food-source for the worm, present on the coverslip at the time of sample preparation. Red arrows indicate the cuticle surrounding the pharyngeal lumen and the buccal cavity, respectively.

We outline three regions-of-interest (ROI) within Fig. 5(a), labelled ①, ②, and ③, that highlight prominent structures within the C. elegans worm. Fig. 5(f) shows a zoom of ROI ①, and qualitatively highlights the features pertaining to the pharynx-bulb, intestinal cavity, and the start of the intestinal lumen, as indicated by the yellow, red, and green arrows, respectively. Fig.

5(j) shows a zoom-in of ROI ②, and depicts fertilized eggs and the distal gonad, as indicated by the yellow and green arrows, respectively. Fig. 5(n) shows a zoom of ROI ③ at the tail-end of the worm and highlights the distal and proximal gonads. To quantitatively visualize the worm's 3D biological RI, we present RBG-colored cross-sectional images of all ROIs at various axial positions. Fig. 5(g,h,i) show quantitative RI cross-sections of ROI ①  at $z = 0.0, +2.4$ and $+4.9$ um, respectively. Fig. 5(k,l,m) and Fig. 5(o,p,q) show quantitative RI cross-sections of ROI ② and ③, respectively, at $z = 0.0, -2.4$ and $-4.9$ um. The bulk tissue of the pharynx and gonads are of relatively consistent RI ~1.35. The fertilized eggs, however, demonstrate high heterogeneity (corroborated by DIC imaging) with RI values ranging from 1.335 to 1.35, and regions with high density of lipid droplets, such as the intestine or the proximal gonads, consist of the most variable scattering. Individual lipid droplets can demonstrate RI values as high as 1.40 and are often directly surrounded by features of low RI ~1.335. Fig. 5(r,s,t) show 3D tomographic visualizations of ROIs ①, ②, and ③, respectively.

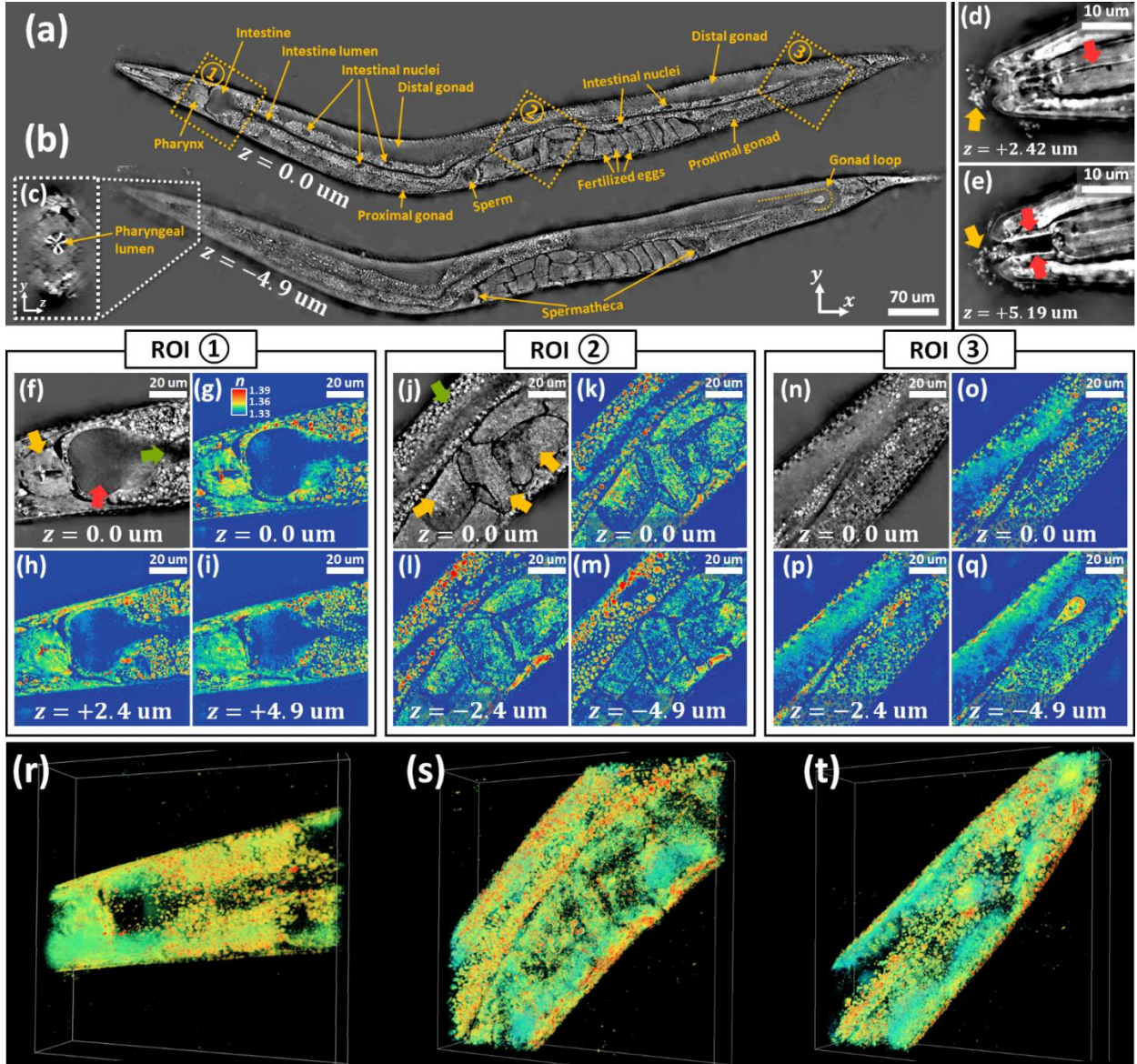

**Fig. 5.** 3D RI reconstruction of a whole *C. elegans* worm is presented. **(a,b)** Lateral cross-sectional planes through the 3D reconstruction volume are shown, at axial positions $z = 0, -4.9$ um, respectively. Major components of the reproductive and digestive systems are labelled. **(c)** Axial cross-sectional plane through the pharynx is shown. **(d,e)** Lateral cross-sectional zooms of the worm's head region at axial positions $z = +2.42, +5.19$ um, respectively, are shown. Zooms of ROIs ①, ②, and ③ are shown in **(f,j,n)**, respectively. Lateral RGB-colored (for easy visual inspection of RI) cross-sections of ROIs ①, ②, and ③, are shown with defocus in **(g,h,i)**, **(k,l,m)**, and **(o,p,q)**, respectively. **(r,s,t)** 3D tomographic renderings of ROIs ①, ②, and ③, respectively, are shown.

## 5. Discussion

This work demonstrates a cost-effective and simple optical hardware system that uses computational imaging for biological imaging problems where no analytical solution exists, as is the case with multiple scattering. This comes at a cost of higher computational requirements than standard ODT. In this work, every angled illumination corresponds to two passes through all layers of the reconstruction volume, which increments the sample's predicted 3D RI one step along its convergence curve. Factors such as increasing the number of acquisitions (to increase the likelihood of correct convergence for the reconstruction) or the resolution of the reconstruction volume further slow down the computation. To address this, future work will leverage the recent advent of big-data processing via GPU acceleration or cloud-computing to significantly reduce reconstruction times.

Another interesting extension would be to explore the maximum resolution achievable using the introduced framework. Previous works have demonstrated that multiple scattering, if considered appropriately, can significantly improve a microscope's resolving power to beyond the diffraction limit [34, 37, 38]. However, this capability for resolution enhancement is difficult to quantify because it depends on the multiple-scattering characteristics of the sample, which are generally unknown. Thus, future work will also explore strategies to estimate the maximum attainable resolution of a sample, based on inferences of its multiple-scattering characteristics.

## 6. Conclusion

In this work, we have introduced a new computational technique that uses the multi-slice beam-propagation model to reconstruct 3D refractive index with high-resolution, even in multiply-scattering samples. Compared to standard ODT techniques, our introduced method offers two key advantages, 1) intensity-only acquisitions enable phase-stable and speckle-free measurements with a dramatically simpler and cheaper optical system. 2) We impose no constraints upon the sample to be weakly scattering, such that suitable samples need not be limited to a sparse distribution of individual cells. This opens up 3D RI microscopy to the more general class of biological samples that include dense cell-clusters or multicellular organisms.

We experimentally demonstrated our method by first reconstructing 3D RI of 3T3 fibroblast cells and C. elegans embryos, as examples of single-cell (weakly-scattering) and multi-cell cluster (multiple-scattering) samples, respectively. In both cases, the MSBP model enabled high-fidelity and robust RI reconstruction. Our final demonstration was the 3D RI reconstruction of a whole C. elegans worm as an example of the applicability of this technique to whole multicellular organisms, even in the presence of multiple-scattering. To the best of our knowledge, this is the first demonstration of high-resolution (NA>1.0) intensity-based 3D RI reconstruction of multiple-scattering biological samples. Due to the simplicity of the introduced optical system, easy hardware additions can be implemented to adapt the system to existing fluorescent microscopes and enable 3D multimodal imaging [47, 52, 53].

**Funding.** Gordon and Betty Moore Foundation's Data-Driven Discovery Initiative (GBMF4562); Ruth L. Kirschstein National Research Service Award (F32GM129966)

**Acknowledgment.** The authors would like to thank Dr. Emrah Bostan and other members of the Computational Imaging Lab for helpful discussions.

# References

# References

# References

# References

# References

1. D. T. Ross, U. Scherf, M. B. Eisen, C. M. Perou, C. Rees, P. Spellman, V. Iyer, S. S. Jeffrey, M. Van de Rijn, and M. Waltham, "Systematic variation in gene expression patterns in human cancer cell lines," Nature genetics **24**, 227-235 (2000).
2. A. Rustom, R. Saffrich, I. Markovic, P. Walther, and H.-H. Gerdes, "Nanotubular highways for intercellular organelle transport," Science **303**, 1007-1010 (2004).
3. M. Okuda, K. Li, M. R. Beard, L. A. Showalter, F. Scholle, S. M. Lemon, and S. A. Weinman, "Mitochondrial injury, oxidative stress, and antioxidant gene expression are induced by hepatitis C virus core protein," Gastroenterology **122**, 366-375 (2002).
4. R. Rizzuto, M. Brini, P. Pizzo, M. Murgia, and T. Pozzan, "Chimeric green fluorescent protein as a tool for visualizing subcellular organelles in living cells," Current biology **5**, 635-642 (1995).
5. W. Choi, C. Fang-Yen, K. Badizadegan, S. Oh, N. Lue, R. R. Dasari, and M. S. Feld, "Tomographic phase microscopy," Nature methods **4**(2007).
6. V. Lauer, "New approach to optical diffraction tomography yielding a vector equation of diffraction tomography and a novel tomographic microscope," Journal of Microscopy **205**, 165-176 (2002).
7. Y. Sung, W. Choi, C. Fang-Yen, K. Badizadegan, R. R. Dasari, and M. S. Feld, "Optical diffraction tomography for high resolution live cell imaging," Optics express **17**, 266-277 (2009).
8. R. Fiolka, K. Wicker, R. Heintzmann, and A. Stemmer, "Simplified approach to diffraction tomography in optical microscopy," Optics express **17**, 12407-12417 (2009).
9. S. Lee, K. Kim, A. Mubarok, A. Panduwirawan, K. Lee, S. Lee, H. Park, and Y. Park, "High-resolution 3-D refractive index tomography and 2-D synthetic aperture imaging of live phytoplankton," Journal of the Optical Society of Korea **18**, 691-697 (2014).
10. T. Kim, R. Zhou, M. Mir, S. D. Babacan, P. S. Carney, L. L. Goddard, and G. Popescu, "White-light diffraction tomography of unlabelled live cells," Nat Photon **8**, 256-263 (2014).
11. T. H. Nguyen, M. E. Kandel, M. Rubessa, M. B. Wheeler, and G. Popescu, "Gradient light interference microscopy for 3D imaging of unlabeled specimens," Nature communications **8**, 210 (2017).
12. Y. Park, M. Diez-Silva, G. Popescu, G. Lykotrafitis, W. Choi, M. S. Feld, and S. Suresh, "Refractive index maps and membrane dynamics of human red blood cells parasitized by Plasmodium falciparum," Proceedings of the National Academy of Sciences **105**, 13730-13735 (2008).
13. W. J. Eldridge, A. Sheinfeld, M. T. Rinehart, and A. Wax, "Imaging deformation of adherent cells due to shear stress using quantitative phase imaging," Optics letters **41**, 352-355 (2016).
14. N. T. Shaked, L. L. Satterwhite, N. Bursac, and A. Wax, "Whole-cell-analysis of live cardiomyocytes using wide-field interferometric phase microscopy," Biomedical optics express **1**, 706-719 (2010).
15. J. Jung, K. Kim, J. Yoon, and Y. Park, "Hyperspectral optical diffraction tomography," Optics express **24**, 2006-2012 (2016).
16. S. S. Kou and C. J. Sheppard, "Image formation in holographic tomography: high-aperture imaging conditions," Applied optics **48**, H168-H175 (2009).
17. M. Habaza, B. Gilboa, Y. Roichman, and N. T. Shaked, "Tomographic phase microscopy with 180 rotation of live cells in suspension by holographic optical tweezers," Optics letters **40**, 1881-1884 (2015).
18. B. Simon, M. Debailleul, M. Houkal, C. Ecoffet, J. Bailleul, J. Lambert, A. Spangenberg, H. Liu, O. Soppera, and O. Haeberlé, "Tomographic diffractive microscopy with isotropic resolution," Optica **4**, 460-463 (2017).
19. A. Devaney, "Inverse-scattering theory within the Rytov approximation," Optics letters **6**, 374-376 (1981).

# Supplementary Information

In this supplement, we describe the experimental preparation protocols used to prepare the biological samples visualized in the main text. We next detail the intuitive basis for the overlapping circles in Fourier spectra that occur when imaging a weakly scattering object, as used in our angle-calibration step. We also describe the procedure used to register and blend together individual reconstructed volumes-of-interest to synthesize a final volume of the whole C. elegans worm. Lastly, we compare the reconstruction fidelities enabled by the multi-slice beam propagation (MSBP) method with those enabled by the 1st Born approximation. This comparison is conducted with the samples presented in the main text, to highlight that the MSBP method is uniquely suited for multiple-scattering biological samples.

## 1. Sample preparation
### A. Preparation of 3T3 fibroblast cells
NIH 3T3 fibroblast cells were cultured in Dulbecco's Modified Eagle Medium (DMEM, Gibco) with 10% fetal bovine serum (FBS; Life Technologies) and 1% penicillin/streptomycin (P/S; Life Technologies) at 37 °C and 5% $CO_2$. For imaging, glass coverslips (12mm diameter, No. 1 thickness; Carolina Biological Supply Co.) were coated with 10μg/mL human fibronectin (Millipore) for 30min at 37 °C. NIH 3T3 cells were passaged onto the coated glass coverslips, cultured for 24hr, and fixed with 3% paraformaldehyde for 20min. Fixed cells were mounted in phosphate buffered saline (PBS; Corning Cellgro) onto a second glass coverslip (24x50mm, No. 1 thickness; Fisher Scientific) and immobilized with sealant (Cytoseal 60; Thermo Scientific).

### B. Preparation of C. Elegans worms and embryos
C. elegans embryos and worms were fixed with 2% formaldehyde for 1 minute, followed by a 10-minute freeze on dry ice. Fisherbrand microscope coverslips (22X40-1.5) were used to prepare the slides. Freeze-cracked slides were washed with PBS buffer twice. Then the samples were mounted within PBS buffer. Worms were generated under normal condition at 20°C. N2 strain was used in this study.

## 2. Symmetrically-shifted circles in intensity spectrum
In the main text, we note that the Fourier amplitude of an intensity measurement of a single scattering sample under angular illumination contains two distinct circles in Fourier space [43]. These two circles are symmetrically positioned in Fourier space around the origin. In our work, this phenomenon is important for two reasons: 1) it is a means to algorithmically calibrate an imaging system for its angular illuminations, and 2) it enables qualitative inspection of how much of a sample's scattering is due to single- vs multiple- scattering. In this section, we aim provide an intuition of where these two circles originate from as well as the relationship of their position to illumination angle.

In the case of weak scattering, the image at the camera plane is formed predominantly by single-scattering interactions within the sample. Thus, the image formation is linear in electric-field and its Fourier spectrum can be described by simple 2D operations. Given a certain operating wavelength (λ) and numerical aperture (NA) for the imaging system, the pupil function of the system is a DC-centered circle with radius of NA/λ in Fourier space. We simulate a noise-only sample with uniform spatial-frequencies, denoted by $x(r)$, where $r$ is the 2D spatial vector. From

Fourier optics, the electric-field at the image plane is given by $y(r) = \mathcal{F}^{-1}\{p(k - k_0) \cdot \mathcal{F}\{x(r)\}\}$, where $k$ is the 2D spatial-frequency vector, $k_0$ is the 2D wave-vector for the illumination angle, $p(k)$ is the system's pupil function, and $\mathcal{F}\{.\} / \mathcal{F}^{-1}\{.\}$ are the Fourier / inverse-Fourier operators [54]. The Fourier-spectra of the electric-field/intensity images are given by $Y(k) = \mathcal{F}\{y(r)\}$ and $I(k) = \mathcal{F}\{|y(r)|^2\}$, respectively.

In Fig. S1 below, we show how illumination angle (described by $k_0$) affects the Fourier spectra of the image-plane's electric field and intensity, $Y(k)$ and $I(k)$, respectively. In all cases where the illumination angle lay within the imaging system's pupil function, $I(k)$ demonstrates a "brightfield" region in Fourier space composed of two symmetrically positioned circular regions [43]. The center-center distance between these two circular regions is set by the illumination angle, $2|k_0|$. Furthermore, $I(k)$ also contains spatial-frequency information in Fourier space surrounding the "brightfield" regions. The visibility of the "brightfield" region compared to the surrounding region is set by the proportion of light that directly transmits through the sample. For example, in the case where the illumination angle lies outside the imaging system's pupil function (i.e., darkfield illumination, Fig. S1(d)), $I(k)$ does not demonstrate the two-circle "brightfield" region. In practice, illumination of a weakly-scattering biological sample results in a strong component of directly transmitted light. Thus, the corresponding $I(k)$ is visually dominated by only the "brightfield" components, as evidenced in the main text Figs. 1(c,e,g). However, in cases of multiple-scattering samples where the directly transmitted light is not the dominant imaging signal, the spatial-frequency signal from outside the "brightfield" regions in Fourier space also become apparent.

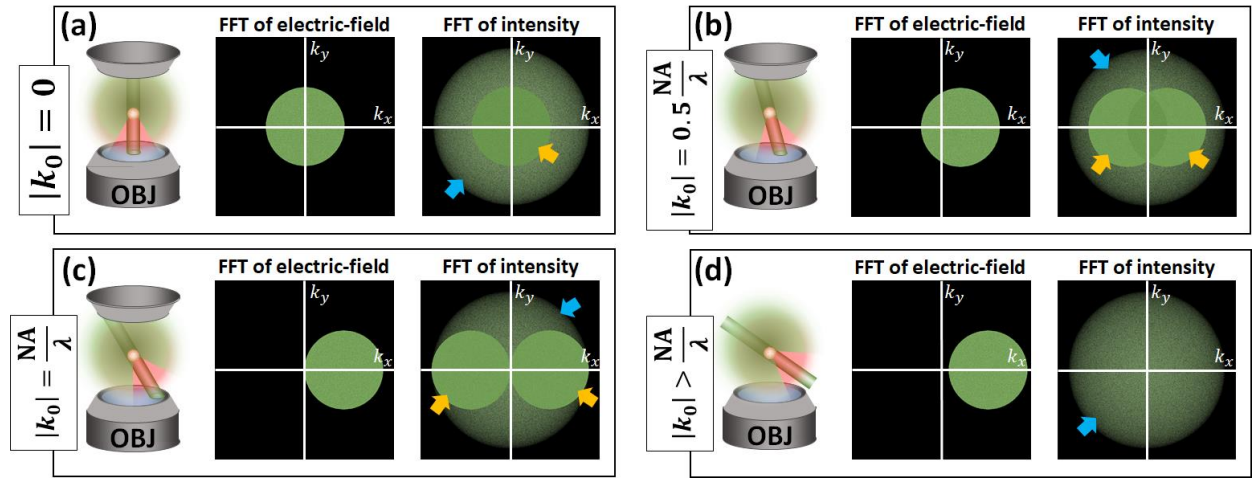

**Fig. S1.** Demonstrating the relationship between the Fourier distributions of the electric-field and intensity measurements, $Y(k)$ and $I(k)$, respectively, for increasing illumination angles, given by wavevectors **(a)** $|k_0| = 0$, **(b)** $|k_0| = 0.5 \, NA/\lambda$, **(c)** $|k_0| = NA/\lambda$, and **(d)** $|k_0| > NA/\lambda$

## 3. Total-volume synthesis from individual 3D reconstructed volumes within C. elegans worm

In the main text, we demonstrated 3D reconstruction of a whole C. elegans worm. Because the worm had a physical length greater than 1 mm, it could not be fully visualized within the imaging objective's FOV (~180 um diameter) – hence, we reconstruct 3D RI within a smaller volume-of-interest (VOI) at a time. Multiple VOIs spanning the whole worm were reconstructed by physically translating the worm. Here, we describe the procedure used to stitch together these separate VOIs into a complete 3D reconstruction of the whole worm.

Fourteen VOIs were required to span the whole C. elegans worm. Each individual VOI had dimensions 1200×1200×100 voxels, and had overlap with adjacent VOIs. The final synthesized RI volume of the whole worm had 1914×10408×118 voxels (2.3 total Gigavoxels). Fig. S2 illustrates the synthesis process of two individual volumes-of-interest (VOIs). This process is repeated for each subsequent VOI to synthesize together the final 3D reconstruction of the whole

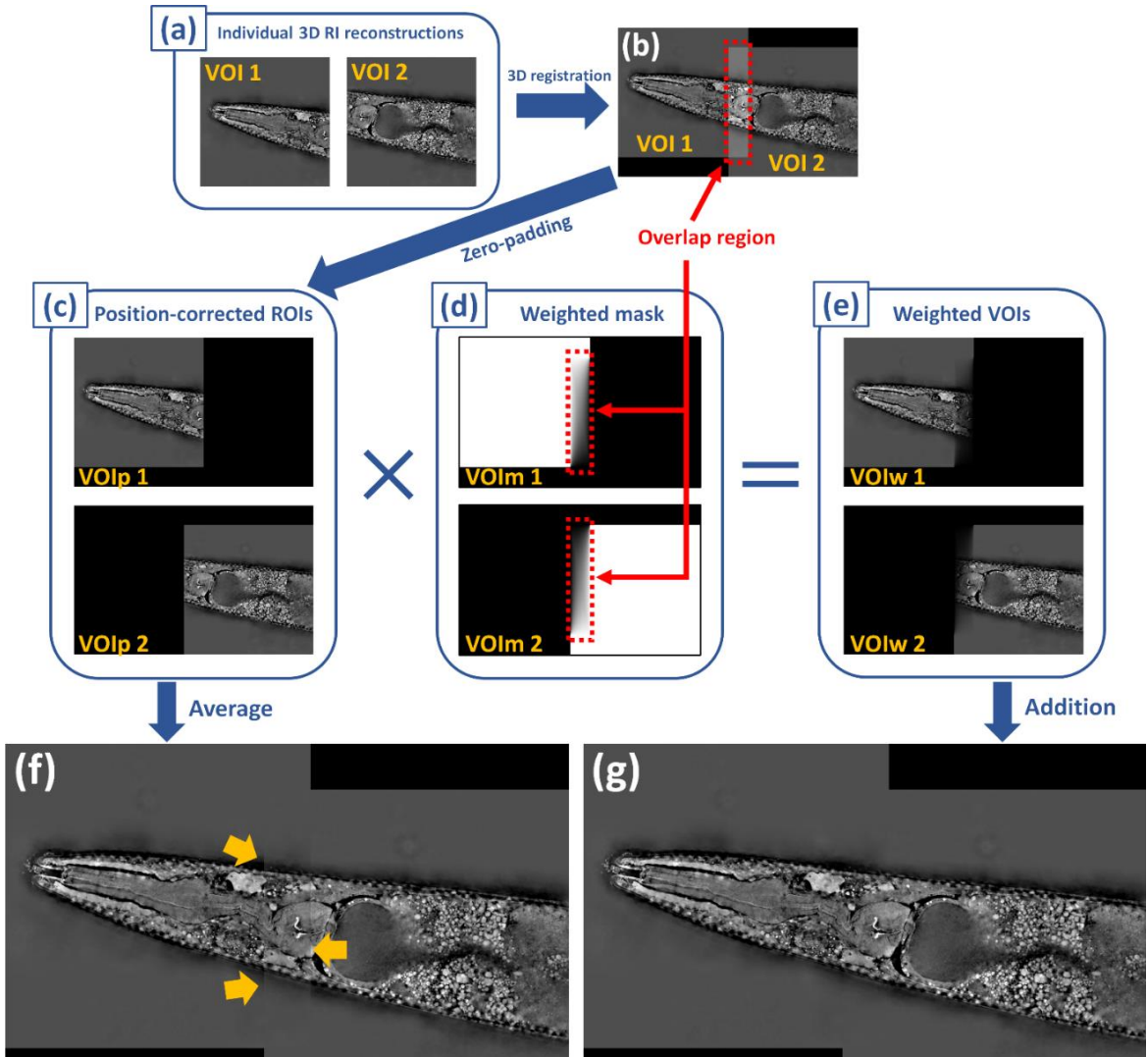

**Fig. S2.** Outline of how two smaller reconstructed RI volumes are synthesized together to form a larger RI volume. **(a)** Start with two adjacent volumes-of-interest (VOIs) that contain a set of the same sample features. **(b)** 3D rigid-body registration algorithms can identify the region of overlap to subsequently calculate the translational difference between VOI1 and VOI2. **(c)** Appropriate zero-padding positions VOI1 and VOI2 properly with respect to each other, in a larger volume space. **(d)** 3D masks are generated. The values within the overlap region (identified in (b)) are set via weighted-average. **(e)** Multiplication of the zero-padded VOIs and the weighted masks results in the weighted VOIs. **(f)** Standard averaging across the zero-padded VOIs results in a synthesized volume with edge artifacts (indicated with yellow arrows). **(g)** However, simple addition across the weighted VOIs results in a synthesize volume with no apparent edge artifacts.

C. elegans worm, shown in the main text Fig. 5.

Fig. S2(a) shows the center depth slice from the first and second VOI. VOI1 captures the head of the worm and VOI2 captures the start of the worm's intestinal tract, with overlap (in this case, the overlap covers the worm's pharynx bulb). Standard 3D rigid-body registration algorithms [55] can identify this overlap and find the 3D translational difference between VOI1 and VOI2 (Fig. S2(b)). This translational difference was used to zero-pad both VOI1 and VOI2 to position them properly with respect to each other, within a greater synthesized volume. We designate the padded versions of VOI1 and VOI2 as VOIp1 and VOIp2 (Fig. S2(b)). At this point, VOIp1 and VOIp2 could in theory be simply averaged over the non-zero values to generate the synthesized volume – however, in practice, such averaging leads to edge-artifacts, as indicated by yellow arrows in Fig. S2(f).

To avoid such artifacts, we adopt a weighted-average scheme for combining VOIp1 and VOIp2. We first generate 3D masks for VOIp1 and VOIp2, designated as VOIm1 and VOIm2, respectively. These masks were initialized with values 1 in locations of all non-zero values in VOIp1 and VOIp2, and zero elsewhere. Afterwards, the values of the masks within the overlap-region between VOI1 and VOI2, as identified earlier, were replaced by a normalized weighted average. This weighted average was calculated based on the distance of each point to the closest edge of the overlap-region consisting of all ones Fig. S2(d). This weighted average protocol is similar to what is often used in standard alpha-blending algorithms [56]. Afterwards, VOIp1 and VOIp2 were multiplied by VOIm1 and VOIm2, to result in the weighted VOIs, VOIw1 and VOIw2, respectively (Fig. S2(e)). A simple addition then results in the fully synthesized volume merging VOI1 and VOI2 (Fig. S2(g)). After comparing Fig. S2(f,g), we see that the weighted average protocol for volume synthesis appropriately avoids the edge artifacts that plague the volume synthesis via direct averaging.

## 3. Reconstruction comparison between 1st Born and multi-slice scattering models

Applying the 1st Born scattering assumption when reconstructing a sample's 3D refractive index (RI) is a popular strategy employed by many diffraction tomography techniques [57]. However, though this assumption is valid for weakly-scattering samples, many biological samples are multiple scattering and cannot be optically characterized by 1st Born scattering. In the main text, we investigated the multi-slice beam propagation (MSBP) model for the robust 3D RI reconstruction of weakly-scattering and multiple scattering biological samples. In Fig. S3, we compare these reconstructions with those obtained via the intensity-based 1st Born scattering inversion model [32]. We note that both models use the same raw data set – the difference between the reconstructions is due purely to which computational model was used.

In Fig. S3(a,b,c), we show an individual raw acquisition from the datasets acquired of the 3T3 fibroblast cell, C. elegans embryo, and C. elegans whole worm samples, respectively. Fig. S3(d,e,f) show their corresponding Fourier spectra. Recall that the 3T3 fibroblast cell is a weakly-scattering sample. This is confirmed by its Fourier spectra, which clearly demonstrates two "brightfield" circles symmetrically positioned in Fourier space. The C. elegans embryo is a multiple-scattering sample. This is also reflected in its Fourier spectrum, which demonstrates significant signal outside the two-circle "brightfield" region (as indicated by yellow arrows in Fig. S3(e)). Finally, the C. elegans worm is also a multiple-scattering sample. Interestingly, its Fourier spectra shows virtually no distinct "brightfield" signal. This signifies that the C. elegans worm is an even stronger multiple-scattering sample than the C. elegans embryo.

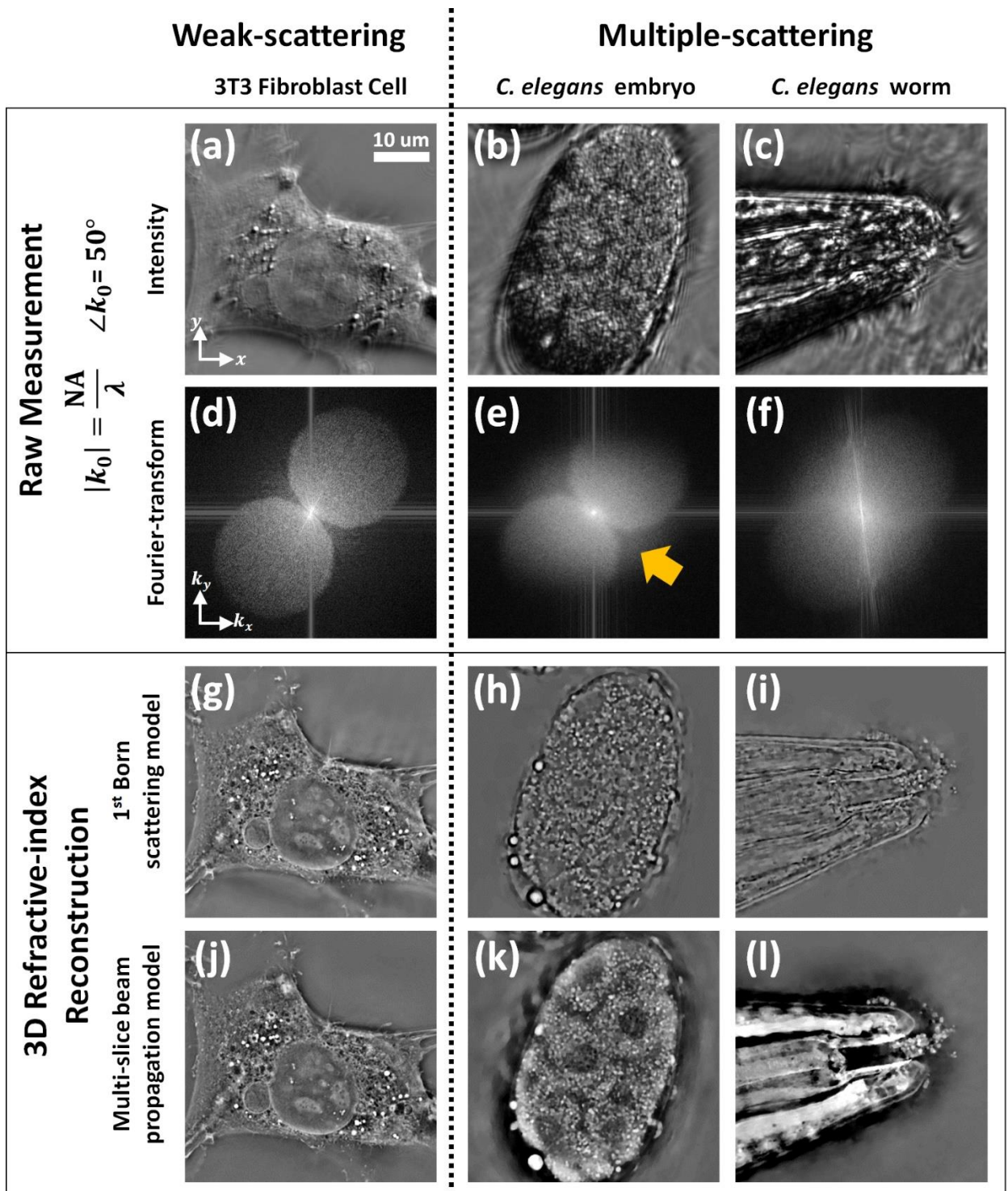

**Fig. S3.** 3D RI reconstruction fidelity between the 1st Born and multi-slice beam propagation (MSBP) models is compared. **(a,b,c)** Example raw acquisitions and **(d,e,f)** associated Fourier spectra are shown from the datasets collected of the 3T3 fibroblast cell, *C. elegans* embryo, and *C. elegans* worm head, respectively. The center cross-sectional plane is shown from the 3D RI volumes of the samples reconstructed via **(g,h,i)** 1st Born and **(j,k,l)** MSBP scattering models.

Fig. S3(g,h,i) and Fig. S3(j,k,l) show the lateral cross-section through the center of the 3D RI volumes of the three samples, as reconstructed using the 1st Born and MSBP inversion models. Because the 3T3 fibroblast cell is a weakly-scattering sample, we expect the 1st Born assumption to be valid and to enable high-fidelity reconstruction. As expected, the reconstructed RI via 1st Born and MSBP match well (Fig. S3(g,j)). In the case of the multiple-scattering C. elegans embryo and worm samples, however, the reconstructions via the 1st Born (Fig. S3(h,i)) and MSBP Fig. S3(k,l)) scattering models show distinct and fundamental differences. 1st Born demonstrates an inability to reconstruct the lower spatial-frequencies in the presence of multiple-scattering, and essentially high-pass filters the RI content. This observation has been affirmed by previous works as well [7]. The degree of high-pass filtering is dependent on the degree of multiple-scattering within the sample, and the 1st Born RI reconstruction of the C. elegans worm (Fig. S3(i)) shows greater high-pass filtering than in that of the embryo (Fig. S3(h)). In both cases, important biological features that are easily visualized in the MSBP reconstruction, such as the individual cells within the embryo and the pharynx muscles and buccal cavity within the worm's head, cannot be visualized in the 1st Born reconstructions.